\documentstyle[12pt,epsf]{article}
\newcommand{\nll}{\nonumber \\}

\newcommand{\bq}{\begin{equation}}
\newcommand{\eq}{\end{equation}}
\newcommand{\ba}{\begin{eqnarray}}
\newcommand{\ea}{\end{eqnarray}}

\newcommand{\nobody}{\rule{0ex}{1ex}}

\newcommand{\vc}{\char'24\hspace{-1ex}c}
\begin{document}
\voffset -2cm
\begin{flushleft}
{DESY 96-111\\
LMU-02/96
}
\end{flushleft}
\begin{center}
\vspace{1.5cm}\hfill\\
{\LARGE \bf{$Z'$ Search in e$^+$e$^-$ Annihilation}}\vspace{1cm}\hfill\\
A. Leike$^{a,}$\footnote{Supported by the German Federal Ministry for
Research and Technology BMBF under contract 05 GMU93P
and by the EC-program CHRX-CT940579}, 
S. Riemann$^b$\\
$\nobody^a$ {\it
Ludwigs--Maximilians-Universit\"at, Sektion Physik, Theresienstra\ss{}e 37,\\
D-80333 M\"unchen, Germany}\\
E-mail: leike@graviton.hep.physik.uni-muenchen.de\\
$\nobody^b$ {\it
DESY - Institut f\"ur Hochenergiephysik, Platanenallee 6, 15738 Zeuthen,
Germany}\\
E-mail: riemanns@ifh.de\\
\end{center}

\begin{abstract}

\noindent
Expectations for
constraints on extra $Z$ bosons are derived for LEP\,2
and future linear $e^+e^-$ colliders.
For typical GUTs, a $Z'$ with  $M_{Z'}\le3~$to$~6  \sqrt{s}$ may cause
observable effects.
The $Z'$ discovery limits are dominated by statistical errors.
However, if a $Z'$ signal is observed, the discrimination between
different  models  becomes much worse  if systematic errors are taken into
account. Discrimination between models is possible for
$M_{Z'} < 3\sqrt{s}$.
A determination of  $Z' f {\bar f}$ couplings independently of models
becomes attractive with future colliders. Anticipated bounds are determined.

 \vspace{1cm}
\end{abstract}
%
\section{Introduction} 

Extra neutral gauge bosons ($Z'$) are predicted in many extensions of the 
Standard Model (SM). 
At future $e^+e^-$ colliders, a $Z'$ can be probed by its virtual effects
on cross sections and asymmetries even if it is much heavier than the
centre-of-mass energy. 
Presently, we have  no experimental indications for  extra neutral gauge
bosons.  
Search results are usually reported as lower limits on the $Z'$
mass, $M_{Z'}^{lim}$, or upper limits on the $ZZ'$ mixing angle for various
$Z'$ models.

In this paper, we continue our study of these limits
started in \cite{lepproc,nlcproc}.
In comparison to \cite{delaguila}, expected systematic errors
are included.
Taking into account radiative and QCD corrections and applying cuts,
we approach a more realistic description of future detectors
and  go beyond \cite{delaguila,zpsari,zpmi}.
In addition to \cite{zpsari,zpmi}, more observables are included.

We set the $ZZ'$ mixing angle equal to zero in accordance with present
experimental constraints  \cite{zmix,zmixl3,zefit}. 
CDF data indicate  that LEP\,2 and LC500 will operate below
a potential $Z'$ peak
\cite{cdfzp94}. 
Similarly, the LHC will be able to detect or exclude
a $Z'$, which could be produced at LC2000 on resonance.
Here, we assume that LC2000 will operate below the $Z'$ peak, too.
Further, we presume universality of generations.
Theories including extra neutral gauge bosons usually predict  new
fermions \cite{maa,e6}. Their effects are neglected here.

We focus on a model-independent approach trying to constrain
the mass and the couplings of  $Z'$ to fermions by different
observables.
For $Z'$ couplings to leptons, this can be done without
further assumptions. A measurement of  $Z'$ couplings to
quarks demands non-zero couplings  to leptons and is
dependent on the latter. 
In addition to the model-independent analysis,
we discuss limits on the $Z'$ mass and couplings for
some typical models given in the Particle Data Book \cite{PDB}.

Neutral currents due to the $Z'$ are
\begin{equation}
\label{zeparm}
J^ \mu_{Z'} = J^\mu_\chi \cos\beta + J^\mu_\psi \sin\beta,\ \ \ 
J^\mu_{Z'} = \alpha_{LR} J^\mu_{3R} - \frac{1}{2\alpha_{LR}} J^\mu_{B-L}.
\end{equation}
Some specified cases are the $\chi,\ \psi$, and
$\eta$ model with
$\beta=0;\pi/2;-\arctan\sqrt{5/3}$ in the $E_6$
GUT\cite{e6,e6lr},
while  special cases discussed in the Left-Right model \cite{e6lr,lr}
are obtained for 
$\alpha_{LR}$ equal to $\sqrt{2/3}$ and $\sqrt{\cot^2{\theta_W}-1}$. 
The first value of $\alpha_{LR}$ reproduces the $\chi$ model while the
second  one gives the Left-Right Symmetric model (LR).
We also consider the Sequential Standard Model (SSM),
where the heavy $Z'$ has exactly the same couplings 
to fermions as the Standard $Z$ boson.

We compare the discovery potential of all relevant reactions in $e^+e^-$
collisions in section~2.
In section~3.1, we discuss model-independent constraints on the
$Z'$ couplings to leptons, which can be derived from the reaction
$e^+e^-\rightarrow f\bar f$.
The model-independent
$Z'$ couplings to quarks are considered in section~3.2.
Section~4 summarizes expected limits for typical
models. We conclude in section~5.

\section{Comparing the final states $f \bar{f}, W^+W^-,$ 4$f$}
%
In this section, we compare the reactions $e^+e^-\rightarrow f\bar f$,
$e^+e^-\rightarrow f_1\bar f_1 f_2\bar f_2$,
and $e^+e^-\rightarrow W^+W^-$ regarding their sensitivity to
indirect $Z'$ signals.
We do not consider  special effects
from the $t$-channel of Bhabha scattering.
For a Born analysis of $e^-e^-\rightarrow e^-e^-$, we refer to \cite{cuypers}.

A (virtual) $Z'$ can be detected by an observable $O$, if it induces
a change $\Delta^{Z'}N$ in the event rate $N_{SM}$, surpassing
the experimental error $\Delta O$, i.e.
\bq
\label{estim1}
\frac{\Delta^{Z'}N}{N_{SM}}\ >\ \frac{\Delta O}{O}.
\eq
For a crude estimate, one can approximate $\Delta^{Z'}N/N_{SM}$
by a ratio of propagators
$D_V=[s-M_V^2+i\Gamma_VM_V]^{-1}$ 
assuming that the $Z'$, the photon and the SM $Z$ boson couple with
similar strengths to SM fermions\footnote{
This is not unreasonable in usual GUTs.}.

We first consider the reaction $e^+e^- \rightarrow f \bar{f}$.
Only the $ZZ'$ interference is important near but off the
$Z$ resonance,
\bq
\label{onres1}
\frac{\Delta^{Z'}N}{N_{SM}} \approx
\frac{|\Re e{D_ZD^{\ast}_{Z'}}|}{|D_Z|^2} =
\frac{s-M_Z^2}{(s-M_{Z'}^2)}.
\eq
Choosing $s=(M_Z+\Gamma_Z/2)^2$, we find from equations (\ref{estim1}) and
(\ref{onres1}) that a $Z'$ with a mass
\bq 
\label{onres}
M_{Z'} > 
M_Z\left( 1+\frac{O}{\Delta O}\frac{4}{5}\frac{\Gamma_Z}{M_Z}\right)^{1/2}
\eq
cannot be excluded.
For $e^+e^-\rightarrow f\bar f$ far off the resonance,
we better consider the $\gamma Z'$ interference. In this case,
the deviation from the Standard Model event rate,
\bq
\frac{\Delta^{Z'}N}{N_{SM}} \approx
\frac{|\Re e{D_\gamma D^{\ast}_{Z'}}|}{|D_\gamma|^2} =
\frac{s}{M_{Z'}^2-s} 
\eq
results to
\bq
\label{offres}
M_{Z'} > \sqrt{s}\left( 1+\frac{O}{\Delta O}\right)^{1/2}.
\eq
For $\Delta O/O = 1\%$, equation (\ref{offres})
leads to a lower bound on the $Z'$ mass,
 $M_{Z'} > M_{Z'}^{lim} \approx 7\sqrt{s}$ (two standard deviations).
Comparing the two expressions 
(\ref{onres}) and  (\ref{offres})
for $Z'$ measurements near  and far off the
$Z$ peak
we see in equation (\ref{onres}) an additional suppression factor $\Gamma_Z/M_Z$.
With  $O/\Delta O \gg 1$ equation (\ref{offres}) is simplified to
$M_{Z'} > \sqrt{s} \times \sqrt{O/\Delta O}$ corresponding to the
 well known scaling law  \cite{zpmi,zepp,rizzo}
\bq
M_{Z'} > \left( s L_{int}\right)^{1/4}
\eq
 with an integrated luminosity $L_{int}$.

Four fermion final states are created in higher order processes. 
Their cross sections are enhanced by resonating $Z$ propagators
near the two--$Z$--boson threshold.
There, the $Z'$ limits are also given by
equation (\ref{onres}). 
As soon as we forbid resonating $Z$ propagators by
invariant mass cuts,
formula (\ref{offres}) should be used. Unfortunately, we are left with
no events in this case.
As a result, four fermion final states will not add any useful
information about a $Z'$.

To get $Z'$ signals in $W$ pair production, one has to assume a
non-zero $Z'WW$ coupling, $g_{Z'WW}=Cg_{ZWW}$. 
Considering the $\gamma Z'$ interference, we get
\bq
\frac{\Delta^{Z'}N}{N_{SM}} \approx
\frac{|\Re e{D_\gamma D^{\ast}_{Z'}}|}{|D_\gamma|^2} =
C\frac{s}{M_{Z'}^2-s} 
\eq
and conclude that a $Z'$ with a mass
\bq
\label{limww}
M_{Z'} < \sqrt{s}\left( 1+C\frac{O}{\Delta O}\right)^{1/2}
\eq
would give a signal in the observable $O$.

The magnitude of $C$ defines the strength of the  $Z'WW$ coupling
and is strongly limited by the decay width of the $Z'$ to $W$
pairs, $\Gamma(Z'\rightarrow W^+W^-)\approx M_{Z'}C^2M_{Z'}^4/M_{Z}^4$.
In a usual GUT,  a reasonable  decay width $\Gamma(Z'\rightarrow W^+W^-)$
results from $C\approx \theta_M\approx M_Z^2/M_{Z'}^2$
where $\theta_M$ is the $ZZ'$ mixing angle.
Taking into account present experimental limits on the $ZZ'$ mixing
and on the $Z'$ mass, we conclude that $C$ must be considerably
smaller than one.
Hence, the limit (\ref{limww}) is always worse than that from fermion
pair production.
The result of these simple estimations (\ref{limww}) is in accordance
with results of \cite{claudio}.

\section{Model-independent $Z'$ Search}
\begin{figure}
\begin{center}
\begin{minipage}[t]{7.8cm}{
\begin{center}
\hspace{-1.7cm}
\mbox{
\epsfysize=7.0cm
\epsffile[0 0 500 500]{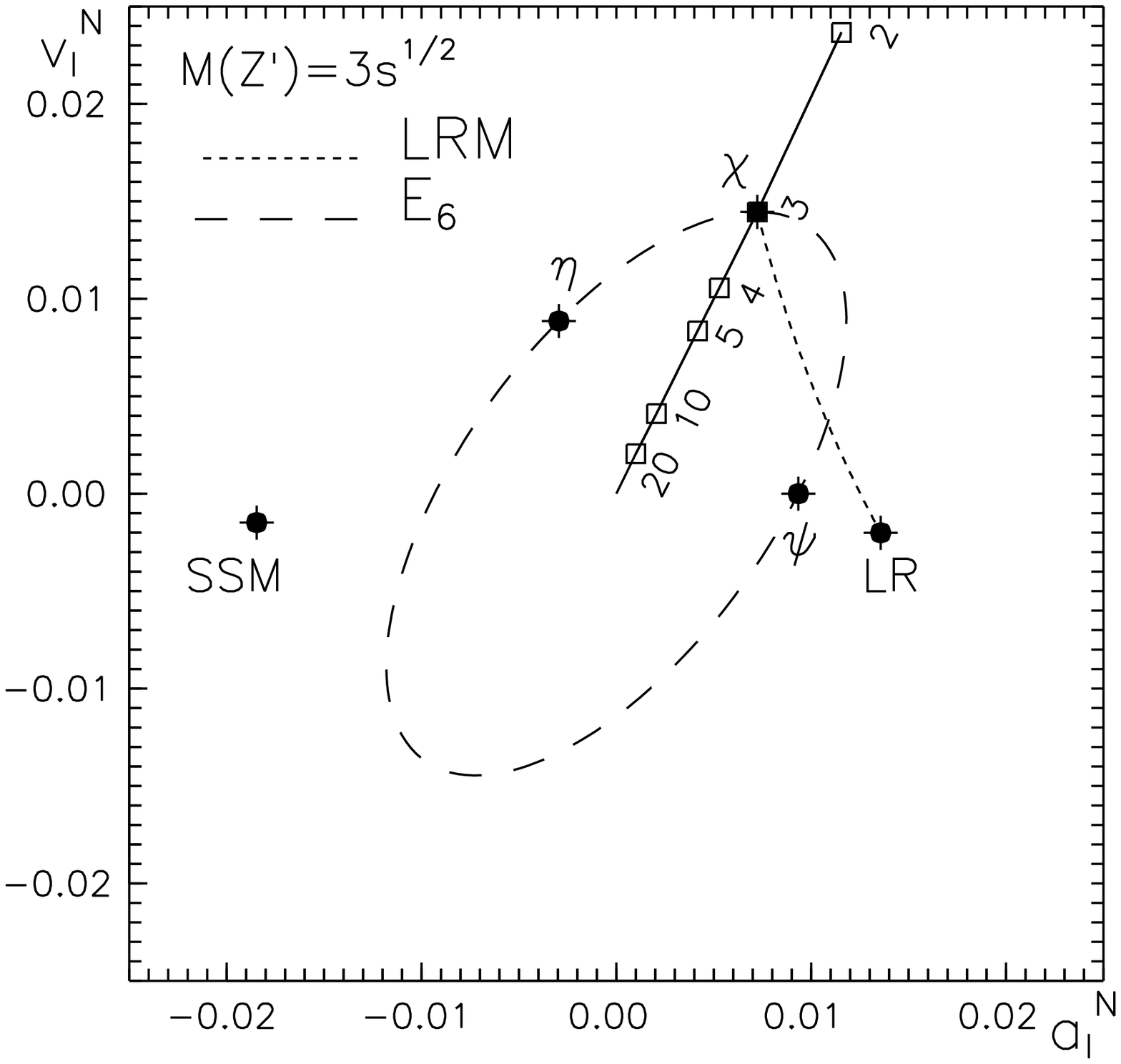}
}
\end{center}
}\end{minipage}
\end{center}
\noindent
{\small\bf Fig.~1: }{\small\it
The normalized vector and axial vector
couplings $Z'l \bar{l}$ for  $M_{Z'} = 3 \sqrt{s}$ in typical GUTs.
For illustration, $M_{Z'}$ for the $\chi$ model is varied
in units of $\sqrt{s}$.
}
\end{figure}

The reaction  $e^+e^-\rightarrow f\bar f$ being most sensitive to a $Z'$
needs further consideration.
We proceed from the following effective Lagrangian,
\begin{equation}
{\cal L} = 
e A_\beta J^\beta_\gamma + g_1 Z_\beta J^\beta_Z + g_2 Z'_\beta J^\beta_{Z'},
\end{equation}
which contains a term describing the additional neutral current interactions 
of the $Z'$ with SM fermions. 
The new interaction leads to an additional amplitude of fermion pair 
production,

\begin{eqnarray}
\label{link}
{\cal M}(Z')  &=& \frac{g_2^2}{s-m_{Z'}^2}
\bar{u}_e\gamma_\beta (\gamma_5 a'_e + v'_e ) u_e \, 
\bar{u}_f\gamma^\beta (\gamma_5 a'_f + v'_f ) u_f
\hspace{5cm} 
\\ \nonumber 
&=&
 -\frac{4 \pi}{s} \left[
\bar{u}_e\gamma_\beta (\gamma_5 a_e^N + v_e^N ) u_e \, 
\bar{u}_f\gamma^\beta (\gamma_5 a_f^N + v_f^N ) u_f \right]
\hspace{5cm} 
\end{eqnarray}
\begin{eqnarray}
\label{link2}
\mbox{\ with\ } 
a_f^N = a'_f \sqrt{\frac{g_2^2}{4 \pi} \frac{s}{m_{Z'}^2-s}}\ ,\ 
v_f^N = v'_f \sqrt{\frac{g_2^2}{4 \pi} \frac{s}{m_{Z'}^2-s}}\ \ 
\mbox{and}\ \ m_{Z'}^2 = M_{Z'}^2-i\Gamma_{Z'} M_{Z'}.
\end{eqnarray}
Fermion pair production is sensitive to
$a_f^N$ and $v_f^N$  and the $Z'$ mass cannot be measured independently of
the $Z'$ couplings.

%
\subsection{$Z'$ couplings to leptons}
For illustrational purposes
the domains of the leptonic $Z'$ couplings  $(a_l^N,\ v_l^N)$
are shown for different models for $M_{Z'}=3\sqrt{s}$ in figure~1.
The variation of the $Z'$ mass for a particular 
$Z'$ model leads to points, which
are on a straight line defined by equation (\ref{link2}).

Measurements of $e^+e^-\rightarrow l\bar l$ ~may constrain  $Z'$ couplings
to leptons based on the observables
\bq
\label{obs}
\sigma_T^l,\ A_{FB}^l,\ A_{LR}^l,\ A_{pol}^\tau
,\ A_{pol,FB}^\tau\mbox{\ and\ } A_{LR,FB}^l.
\eq
The index $l$ stands for electrons and muons in the
final state (only the $s$ channel is considered for electrons).
Neglecting fermion masses, we have the following relations in
Born approximation (assuming lepton universality):
\bq
\label{bornobs}
A_{LR}^l = A_{pol}^\tau = \frac{4}{3} A_{pol,FB}^\tau =\frac{4}{3}A_{LR,FB}^l.
\eq
All observables in equation (\ref{bornobs}) depend on the same combination
of $Z'$ couplings to leptons. We will concentrate on
$A_{FB}^l$ and  $\sigma_T^l$ combine it with  $A_{LR}^l$ if the
beams are polarized or with $A_{pol}^\tau$ without polarized beams.

We calculate the Standard Model  predictions of all
available  observables,
$O_i($SM$)$, and compare them with the predictions,
$O_i($SM$,v_l^N,a_l^N)$, in a theory including a $Z'$.
We define
\bq
\label{chisq}
\chi^2 = \sum_{i}
\left[\frac{O_i(\mbox{SM})-O_i(\mbox{SM},v_l^N,a_l^N)}{\Delta O_i}
\right]^2 ,
\eq
where $\Delta O_i$ are experimental errors.
For $\chi^2>\chi^2_{min}+5.99$, the values for the parameters
$(a_l^N,v_l^N)$ are excluded at  95\% confidence level.

Simple approximate formulae for the excluded regions in the  $(a_l^N,v_l^N)$ 
plane can be obtained in the Born approximation,
\begin{eqnarray}
\begin{array}{rllcccc}
\sigma_T^l& \mbox{ detects\ a\ Z}'\ \mbox{if} &
 \displaystyle{\left(\frac{v_l^N}{H_v}\right)^2
+\left(\frac{a_l^N}{H_a}\right)^2} &\ge& 1,&
 H_{v,a}\sim &\sqrt{\Delta\sigma_T^l/\sigma_T^l} \\   \\
A_{FB}^l &\mbox{ detects\ a\ Z}'\ \mbox{if} &
 \displaystyle{\left|\left(\frac{v_l^N}{H'_v}\right)^2
-\left(\frac{a_l^N}{H'_a}\right)^2
\right|} &\ge& 1,&
H'_{v,a}\sim& \sqrt{\Delta A_{FB}^l}  \\   \\
A_{LR}^l &\mbox{ detects\ a\ Z}'\ \mbox{if} &
 \displaystyle{\left|\left(\frac{v_l^N}{H''_v}\right)
\left(\frac{a_l^N}{H''_a}\right)\right|}
&\ge& 1,&   H''_{a,v}\sim  &\sqrt{\Delta A_{LR}^l}.
\label{bornlim}
\end{array}
\ea
Note that the axes of the ellipse $H_{v,a}$ and the hyperbolas
$H_{v,a}^{(}\!\!\!\nobody '\nobody^{,}\nobody ''\nobody^{)}\ $ 
do not depend on the $Z'$ model \cite{zpmi}.
This offers the interesting simpler
possibility of a model-independent 2-parameter analysis.

Below the $Z'$ peak, the $Z'$ can be detected through small deviations
of observables from their SM predictions.
Therefore, radiative corrections have to be included to meet the expected 
experimental precision with accurate theoretical predictions.
Due to the radiative return to the $Z$ resonance
the energy spectrum of the radiated photons is peaking around
$E_\gamma/E_{beam}\approx 1-M_Z^2/s$.
Events with such hard photons  ``pollute''
the potential signal resulting in much weaker $Z'$ limits than predicted in
the Born approximation. 
Therefore, they should be eliminated from  a $Z'$ search
by a cut on the
photon energy, $\Delta = E_\gamma/E_{beam}< 1-M_Z^2/s$ or by  cuts
on the acollinearity angle and the energy of the outgoing fermions.

Our analysis is performed with the Fortran program {\tt ZEFIT}
\cite{zefit_code}, which has to be used together with  {\tt ZFITTER}
\cite{zfitter,rcorr}.
Hence, we take into account  all SM corrections and all possibilities
to apply kinematical cuts available in {\tt ZFITTER}. 
{\tt ZEFIT} contains the additional $Z'$ contributions.
For the present studies, we
adapted {\tt ZEFIT} to a model-independent $Z'$ analysis.
QED corrections to the $Z'$ contributions are applied  \cite{slr}.

The following scenarios of $e^+e^-$ colliders are considered:\vspace{1ex}

\begin{tabular}{llll}
\label{colliders}
LEP\,2  & $\sqrt{s}=190$\,GeV & $L_{int}=500pb^{-1}$ &
no polarization, 4 experiments \\
LEP\,2P & $\sqrt{s}=190$\,GeV & $L_{int}=500pb^{-1}$ &
P=80\% $e^-$ polarization, 4 experiments \\
LC500  & $\sqrt{s}=500$\,GeV & $L_{int}=20fb^{-1}$ &
P=80\% $e^-$ polarization\\ 
LC2000 & $\sqrt{s}=2$\,TeV & $L_{int}=320fb^{-1}$ &
P=80\% $e^-$ polarization
\end{tabular}\vspace{1ex}

The statistical errors for $N$ detected events are
\begin{equation}
\label{stat}
\frac{\Delta\sigma_T}{\sigma_T}=\frac{1}{\sqrt{N}},\ \ \
\Delta A_{FB} = \Delta A_{pol} = \Delta A = \sqrt{\frac{1-A^2}{N}},\ \ \ 
\Delta A_{LR} = \sqrt{ \frac{1 -(P A_{LR})^2}{N P^2} }\ .
\end{equation}

\begin{figure}
\begin{center}
\begin{center}
\hspace{-1.7cm}
\mbox{
\epsfxsize=7.8cm
\epsffile{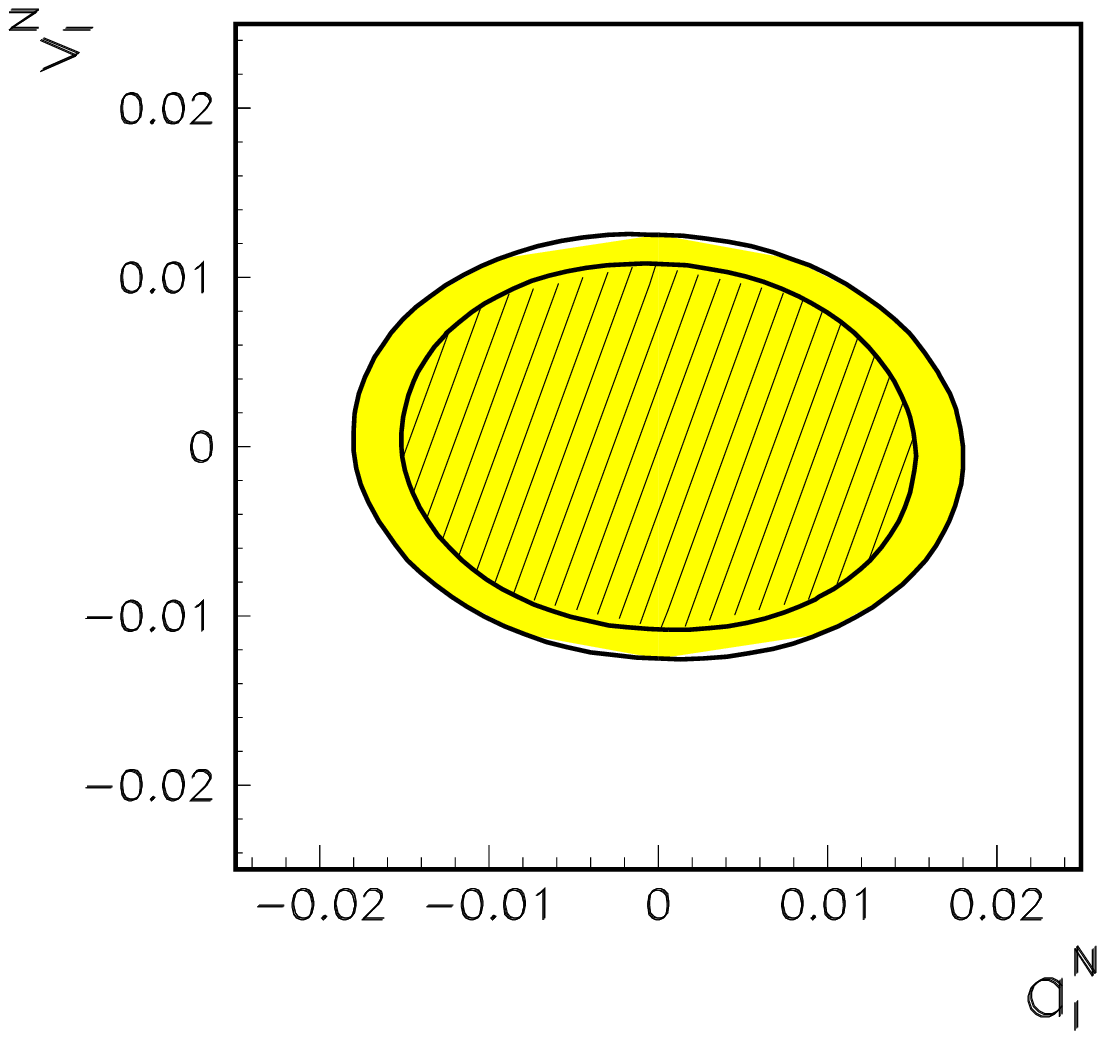}
}
\end{center}
\end{center}
\noindent
{\small\bf Fig.~2: }{\small\it
Expectations for allowed regions for
normalized $Z'l \bar{l}$ couplings (95\% CL) derived from measuring
$\sigma_T^l$ at LEP\,2
with (shaded) and without (hatched) systematic errors.
Radiative corrections are included.
}
\end{figure}
%
We assume a systematic luminosity error of 0.5\% .
Further, we include a systematic error of 0.5\% for the measurement of each
observable. The error of $A_{pol}^\tau$ is estimated with 5\%.

We avoid hard photons by applying a cut on the photon energy and take
 $\Delta=0.7$ for LEP\,2, $\Delta=0.9$ for LC500 and
$\Delta=0.98$ for LC2000.
As a simple simulation of the detector acceptance, 
we demand that the angle between the outgoing
leptons and the beam axis is larger than $20^\circ$. 

The  errors for all considered observables are then
\begin{equation}
\begin{array}{rrclrclrclrcl}
\label{comberr}
LEP2P: &
\Delta\sigma_T^l/\sigma_T^l&=&1.1\%,\ \
\Delta A_{FB}^l&=&1.1\%,\ \
\Delta A_{LR}^l&=&1.3\% \ \
\Delta A_{pol}^\tau&=&5\%,\nll
LC: &
\Delta\sigma_T^l/\sigma_T^l&=&1\%,\ \
\Delta A_{FB}^l&=&1\%,\ \
\Delta A_{LR}^l&=&1.2\% \ \
\Delta A_{pol}^\tau&=&5\%
\end{array}
\end{equation}
The collider parameters for the two LC scenarios are chosen such that
the observables have the same relative errors.
Possible correlations between the errors of different observables 
are neglected.

Values of $a_l^N, v_l^N$, which cannot be excluded  with 95\% confidence
by a measurement of
the total cross section,  $\sigma_T^l$,
are shown in  figure~2.
The bounds on the normalized couplings become roughly 15\% narrower
if systematic errors are neglected.
Analogue considerations with the other observables in (\ref{obs}) give
similar results.
If the radiative return to the Z peak is prevented the $Z' l{\bar l}$ bounds are
almost indistinguishable from  those obtained in the Born approximation.
The weak corrections contained in $\sigma_T^l(SM)$ are dropped out
in our predictions calculating $\chi^2$ as proposed in eq. (\ref{chisq}).
However,  radiative and weak corrections to observables are large
enough to forbid the usage of Born formulae in fits to real data. 
For illustration, in table~1 the expected cross sections $\sigma_T$($e^+e^-
\rightarrow \mu \bar{\mu}$) at $\sqrt{s}$=190~GeV and $\sqrt{s}$=500~GeV are listed
in Born approximation, with a cut on the photon energy and with
cuts on both photon energy and angular acceptance.
The values correspond to the predictions of the SM and
the $\chi$ model with a $Z'$ mass $m_{Z'}$=1~TeV and $m_{Z'}$=2~TeV.

\begin{table}[htb]
\begin{center}
\begin{tabular}{lrrrr}\hline\\
 &  $\sqrt{s}$ [GeV]&~~~~~~~~~~~~~Born~~&~~~~~~~~~Born+RC~~&Born+RC+ang.cut\\
 &                        &           &               &                \\ \hline
        &                &           &               &                \\
  SM    &       190      &    3 379  &        3 753  &         3 382   \\
        &        500     &      465  &          559  &           509   \\
        &                &           &               &                 \\
 $\chi$ model    & 190   &    3 329  &        3 677  &         3 338   \\
$m_{Z'}$=1000 GeV& 500   &      405  &          499  &           449   \\
                         &           &               &                 \\
 $ \chi$ model   & 190   &    3 367  &        3 713  &         3 372   \\
$m_{Z'}$=2000 GeV& 500   &      453  &          547  &           492   \\
        &                &           &               &                \\
\hline
\end{tabular}\medskip
\end{center}
{\small \bf Table~1: }{\small\it
Cross sections in fb  for  $e^+e^-
\rightarrow \mu \bar{\mu}$ in Born approximation, 
including radiative corrections (RC)
with cut on the energy of photons
emitted in the initial state ($\Delta$=0.7 if $\sqrt{s}$=190~GeV and
$\Delta$=0.9 if $\sqrt{s}$=500~GeV) and with radiative corrections with
cuts on both, on $\Delta$ and on angular
acceptance ($\theta \ge$20$^{\circ}$).
} \end{table}
%

Figure~3a contains model-independent limits on
$Z'$ couplings to leptons as they may be
expected from experiments at LEP2/LEP2P. They are  shrinked when more
observables are considered.
Radiative corrections are taken into account.
$\sigma_T^l$ constrains both the vector and axial vector couplings
while $A_{FB}^l$ restricts mainly axial vector couplings.
$A_{LR}$ reduces the allowed regions only slightly.
The excluded regions agree reasonably with estimates using the Born
formulae (\ref{bornlim}).
Thus, if errors different from our assumptions (\ref{comberr}) are preferred
these formulae can be used to estimate the shift of the bounds in the
$(a_l^N,v_l^N)$ plane.
A comparison of figure~3a with figure~1 shows,
which observable is important to constrain a particular $Z'$ model.

If all leptonic observables are combined  the allowed area in the
$(a_l^N,v_l^N)$ plane will be reduced as is
demonstrated in figure~3b for LEP2 and LC500.
These two regions are  indistinguishable from those of
LEP\,2P and LC2000, respectively.

\begin{figure}
\begin{minipage}[t]{7.8cm}{
\begin{center}
\hspace{-1.7cm}
\mbox{
\epsfysize=7.8cm
\epsffile{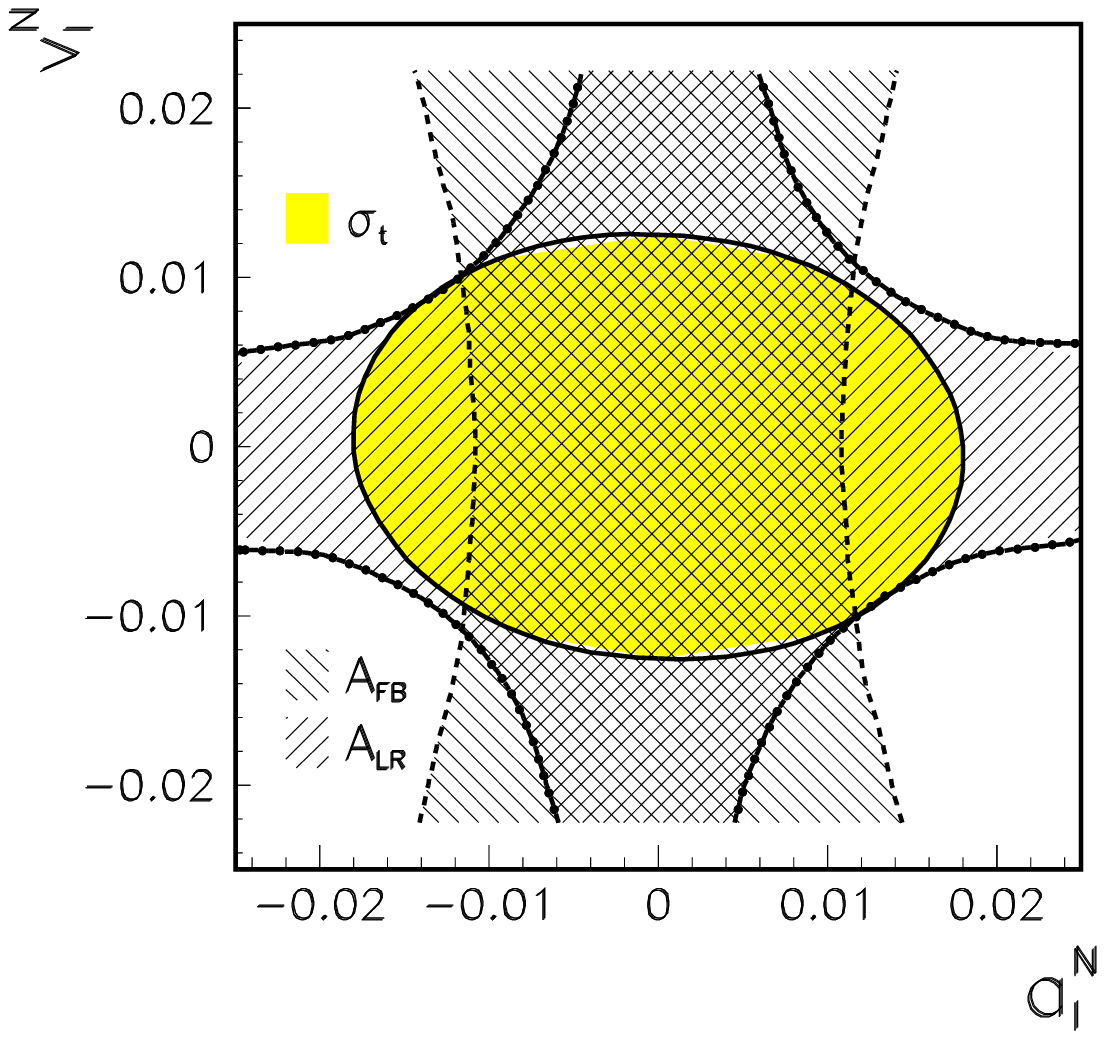}
}
\end{center}
\noindent
{\small\bf Fig.~3a: }{\small\it
Areas of $(a_l^N,\ v_l^N)$ values for which the extended gauge theory's
predictions are indistinguishable from the SM (95\% CL) at LEP\,2P.
Models inside the  ellipse cannot be detected with $\sigma_T^l$
measurements.
Models inside the hatched areas with 
falling (rising) lines cannot be resolved with $A_{FB}^l$ ($A_{LR}^l$).
}
}\end{minipage}
\hspace{0.5cm}
\begin{minipage}[t]{7.8cm}{
\begin{center}
\hspace{-1.7cm}
\mbox{
\epsfysize=7.8cm
\epsffile{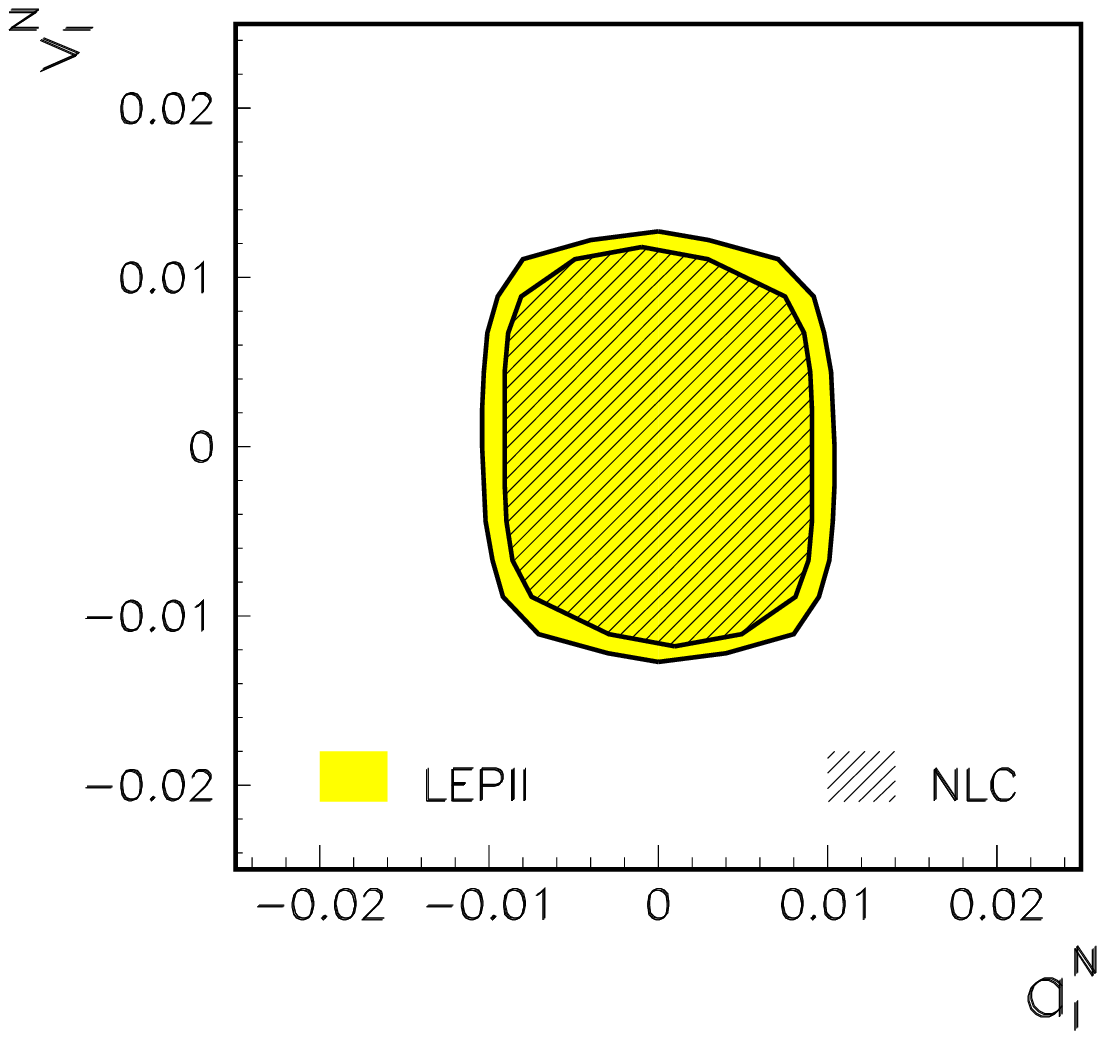}
}
\end{center}
\noindent {\small\bf Fig.~3b: }{\small\it 
Areas of $(a_l^N,\ v_l^N)$ values  for which the extended gauge theory's
predictions are indistinguishable from the SM (95\% CL)
at different colliders based on  all leptonic observables.
}
}\end{minipage}
\end{figure}
\vspace*{0.5cm}
%

\begin{figure}
\begin{minipage}[t]{7.8cm}{
\begin{center}
\hspace{-1.7cm}
\mbox{
\epsfysize=7.8cm
\epsffile{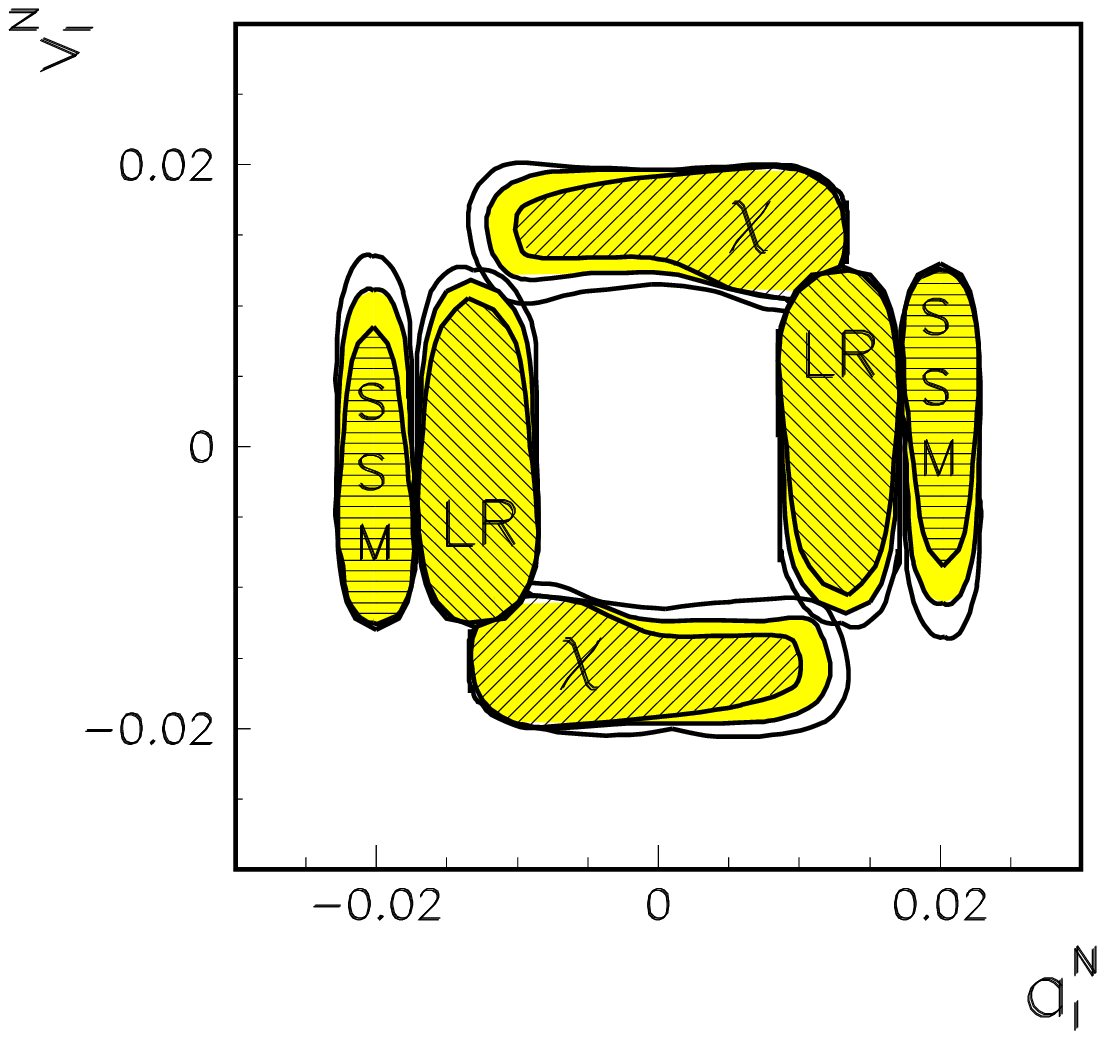}
}
\end{center}
\noindent
{\small\bf Fig.~4a: }{\small\it
Resolution power of LEP2
(95\% CL) based on a combination of all leptonic observables
for $M_{Z'}$=550~GeV.
Different models cannot be resolved with 95\% CL within the hatched 
(shaded) areas if $\Delta A_{pol}^\tau$=3.5\%(5\%).
White areas result from $\sigma_T$ and $A_{FB}$ only.
}
}\end{minipage}
\hspace{0.5cm}
\begin{minipage}[t]{7.8cm}{
\begin{center}
\hspace{-1.7cm}
\mbox{
\epsfysize=7.8cm
\epsffile{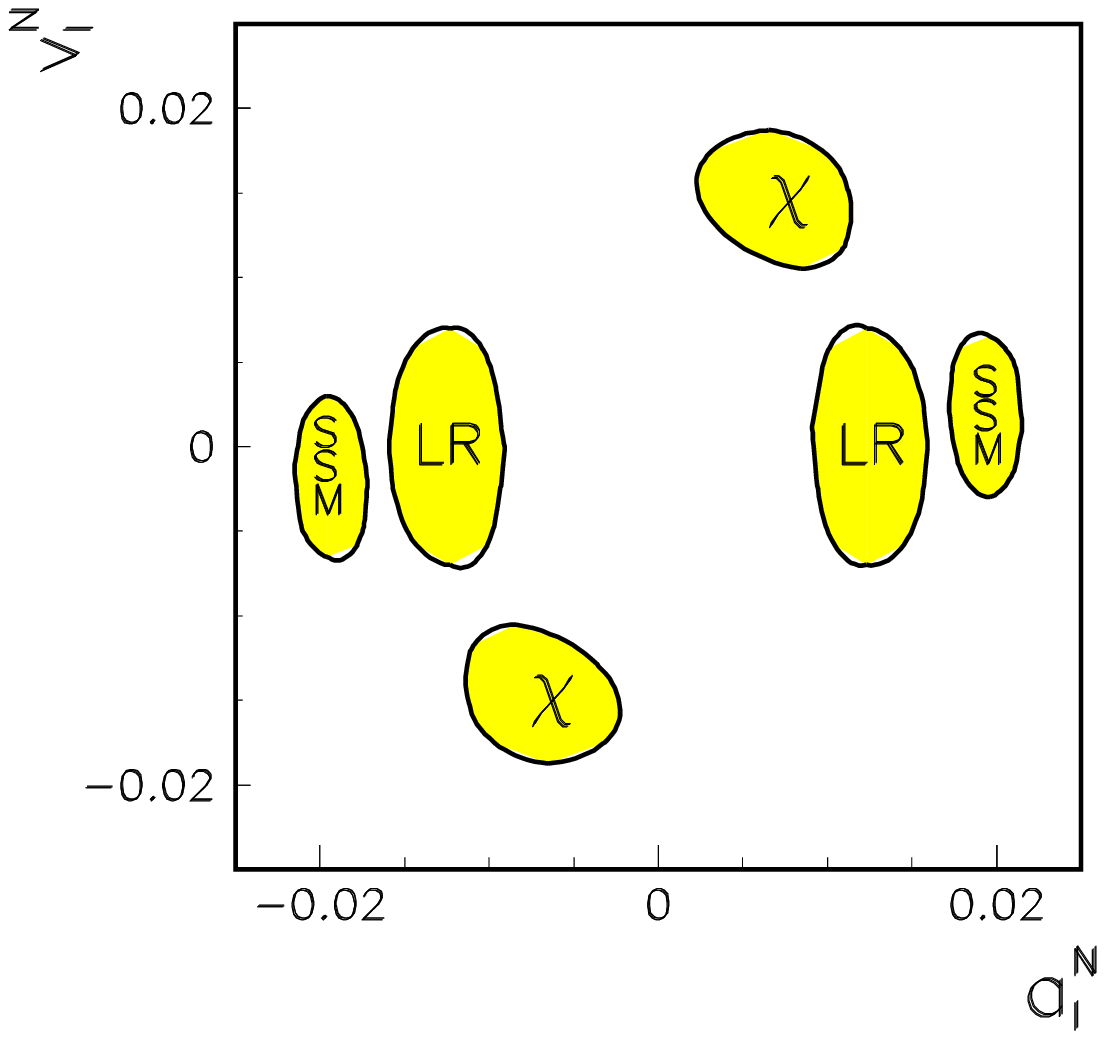}
}
\end{center}
\noindent
{\small\bf Fig.~4b: }{\small\it
Resolution power of LC500
(95\% CL) for different models and $M_{Z'}=1.5\,TeV$
based on a combination of all leptonic observables.
}
}\end{minipage}
\end{figure}
%
\begin{figure}
\begin{center}
\begin{minipage}[t]{7.8cm}{
\begin{center}
\hspace{-1.7cm}
\mbox{
\epsfysize=7.8cm
\epsffile{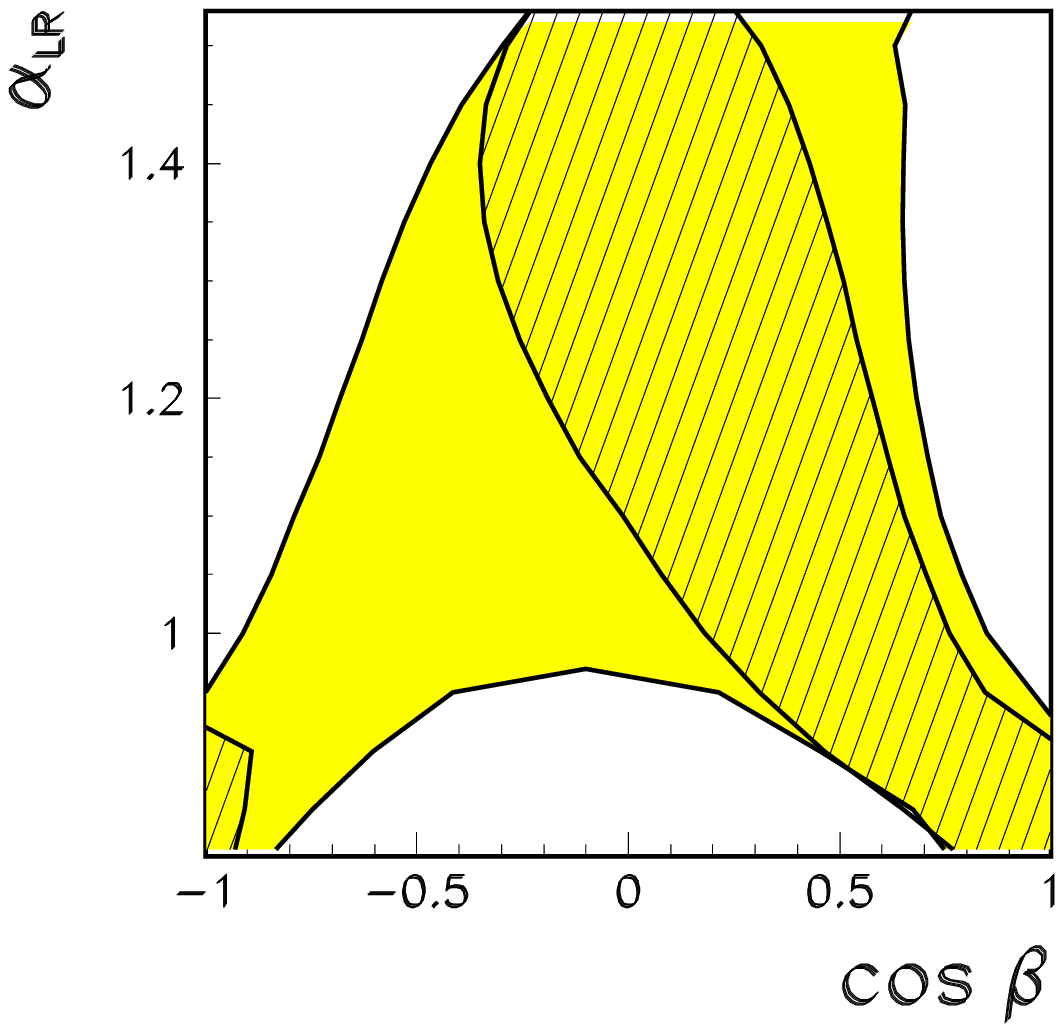}
}
\end{center}
}\end{minipage}
\end{center}
\noindent
{\small\bf Fig.~4c: }{\small\it
Confusion regions between $E_6$ and left-right models at LC500
for $M_{Z'}=1.5\,TeV$ based on $\sigma_T^l$ and $A_{FB}^l$ (shaded)
and on $\sigma_T^l,\ A_{FB}^l$ and $A_{LR}^l$ (hatched).
}
\end{figure}

If a $Z'$ with $M_{Z'}=550\,GeV$ would be found at Tevatron soon,
a model identification may be tried at LEP\,2.
We illustrate this in figure~4a. Typically,
the allowed regions in the  $(a_l^N,\ v_l^N)$ plane  are clearly
off the point (0,0).
In contrast to $\sigma_T^l$ and $A_{FB}^l$, the
polarisation asymmetries $A_{LR}$,
$A_{pol}^\tau$ are sensitive to the sign of the $Z'$ couplings.
Even a measurement of $A_{pol}^\tau$ 
with relatively large errors could
help to reduce sign ambiguities.
Polarized beams also would remove these ambiguities as shown in figure~4b.
In experiments at LC500, the three models SSM, LR and $\chi$  can be
distinguished for $M_{Z'}=3\sqrt{s}$.
Note that the simultaneous change of signs of both leptonic $Z'$ couplings
can never be detected by the reaction $e^+e^-\rightarrow l\bar l$.

A qualitative discrimination between E$_6$ and LR models can be done
by a superposition of figures~1 and 4b.
A one-parameter fit has to be performed for quantitative estimates.
In  figure 4c, we
assume a $Z'$ with a mass of  $M_{Z'}=1.5 $TeV and derive the
region of confusion for the model parameters $\cos \beta$ and $\alpha_{LR}$
based on a measurement of leptonic observables.
If, e.g., a $Z'$ with $M_{Z'}=1.5 $TeV occurs in the
$\chi$ model,
the region $-22^\circ < \beta < 40^\circ$ in the $E_6$ GUT
cannot be distinguished from Left-Right models with
$\alpha_{LR}<0.91$ by leptonic observables. Remember that 
the $\chi$ model corresponds to $\beta = 0$ and
to $\alpha_{LR}=\sqrt{2/3}$. 

Finally, we should note that experimental
restrictions to $a_l^N$ and $v_l^N$ by the three leptonic observables
could lead to contradicting results.
This would be an indication for new physics not related to a $Z'$.

%
\subsection{$Z'$ couplings to quarks}

We now perform a model-independent analysis of $Z'q\bar q$ couplings.
Hadronic observables depend on  $Z'q\bar q$ couplings {\it and}
$Z'l\bar l$ couplings and can be measured with a good accuracy
for all assumed collider scenarios.
All $Z'q \bar q$ couplings contribute combined to the
hadronic observables
\ba 
\label{mdobs}
LEP2P && R^{had}=\frac{\sigma_T^{had}}{\sigma_T^\mu}
=\frac{\sigma_T^{u+d+s+c+b}}{\sigma_T^\mu},
\ \Delta R^{had}=1.0\%,\ \ 
A_{LR}^{had}=A_{LR}^{u+d+s+c+b},\ \Delta A_{LR}^{had}=0.8\%,\nll
LC & & R^{had}=\frac{\sigma_T^{had}}{\sigma_T^\mu}
=\frac{\sigma_T^{u+d+s+c+b}}{\sigma_T^\mu},
\ \Delta R^{had}=0.9\%,\ \ 
A_{LR}^{had}=A_{LR}^{u+d+s+c+b},\ \Delta A_{LR}^{had}=0.7\%.\nll
 && 
\ea
In order to pick up single flavours one needs advanced techniques of flavour
identification.
From the experience of LEP\,1 and SLD one expects relatively
small errors of $b$-quark and $c$-quark observables at future $e^+ e^-$ colliders.
Here, we restrict our studies to  $Z'b\bar b$ couplings only.
We do not apply angular restrictions to outgoing quarks.
Taking into account inefficiencies and systematic errors of flavour tagging
we include the following observables and their experimental uncertainties:
\ba 
LEP2P & R_b=\displaystyle{
\frac{\sigma_T^b}{\sigma_T^{had}}}\ \Delta R_b=2.5\%,\ \
A_{FB}^b,\ \Delta A_{FB}^b=2.2\%,\ \  
A_{LR}^b,\ \Delta A_{LR}^b=1.5\%,\nll
LC & R_b=\displaystyle{
\frac{\sigma_T^b}{\sigma_T^{had}}},\ \Delta R_b=2.2\%,\ \
A_{FB}^b,\ \Delta A_{FB}^b=2.0\%,\ \  
A_{LR}^b,\ \Delta A_{LR}^b=1.5\%.
\ea

%
\begin{figure}[t]
\begin{center}
\begin{minipage}[t]{7.8cm} {
\begin{center}
\hspace{-1.7cm} \mbox{ \epsfxsize=7.0cm \epsffile[0 0 500 500]{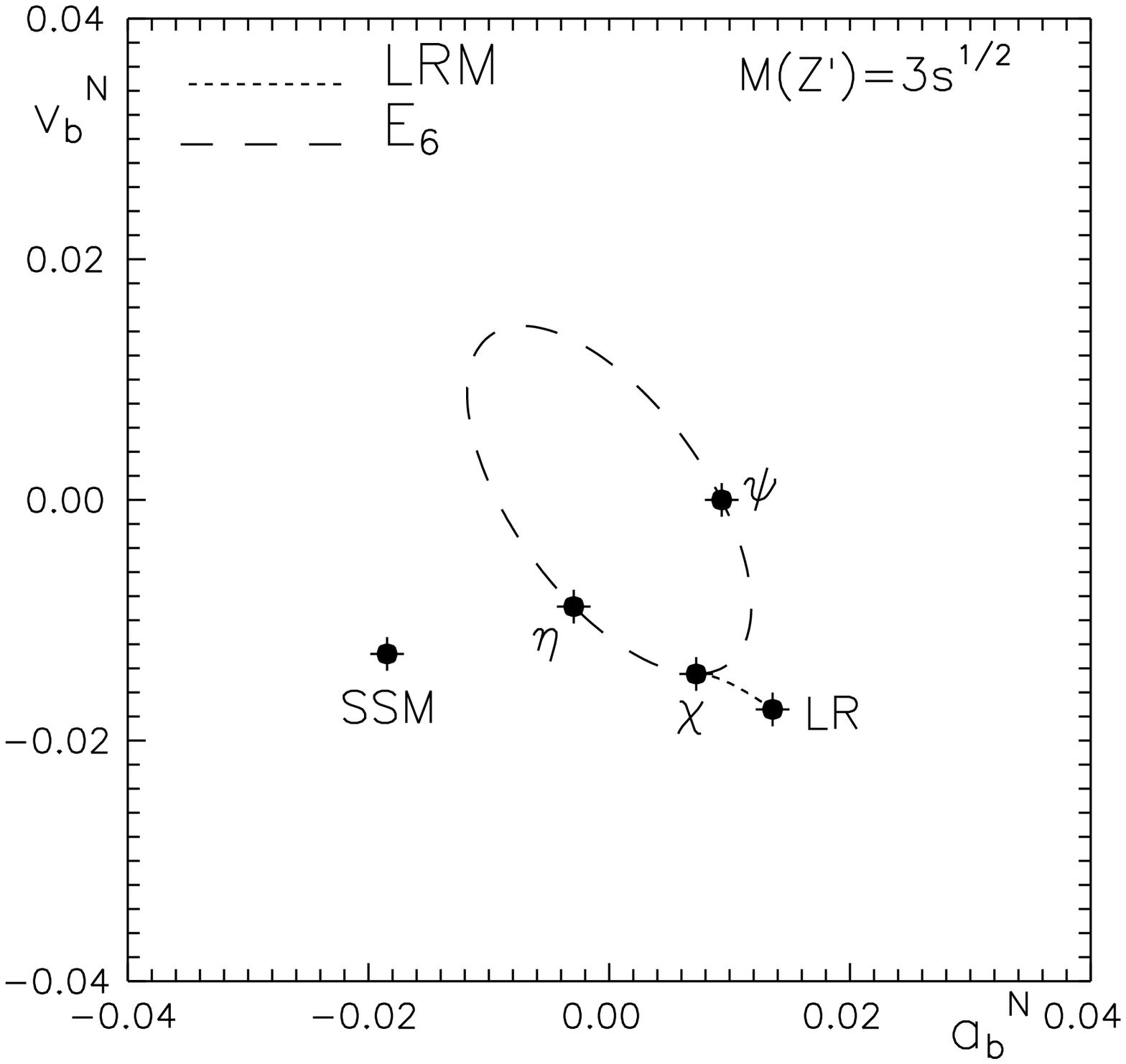}
}
\end{center}
}\end{minipage}
\end{center}
\noindent {\small\bf Fig.~5: }{\small\it Normalized $Z'b\bar{b}$ couplings
for different models with $M_{Z'} = 3 \sqrt{s}$.
}
\end{figure}
%

Figure~5 shows 
the normalized couplings $a_b^N, v_b^N$ for different $Z'$ models
if $M_{Z'}=3\sqrt{s}$.
The $Z'b\bar b$ couplings are not very sensitive to the parameter $\alpha_{LR}$
of Left-Right models whereas the variation of the parameter
$\beta$ ($E_6$ GUTs) leads to significantly different $Z'b\bar b$ couplings.
Especially in the SSM the $Z'b\bar b$ coupling  is rather large 
and quite different from the other models considered.

Non-vanishing  leptonic $Z'$ couplings are a necessary ingredient in order
to get any information about the $Z'$ couplings to quarks from experiments at
$e^+ e^-$ colliders.
Assuming a $Z'$ with leptonic couplings at the boundary of figure~3b,
one is unable to exclude $Z'b\bar b$ couplings inside the marked  areas
shown in figure~6a for  LEP\,2 and  LEP\,2P.
Different observables shrink various regions.
Figure~6a emphasizes that polarized beams give a large improvement to
the measurement of $Z'b\bar b$ couplings. 
Of course, the allowed area for $a_b^N, v_b^N$ couplings
depends on the choice of the $Z'$ couplings to the initial state.
Figure 6b shows the corresponding resolution power
for $Z'b{\bar b}$ couplings  expected with  an LC500.
This region contains the point
$(a_b^N,v_b^N)=(0,0)$.

%
\begin{figure}
\begin{minipage}[t]{7.8cm}{
\begin{center}
\mbox{ \epsfysize=7.8cm \epsffile{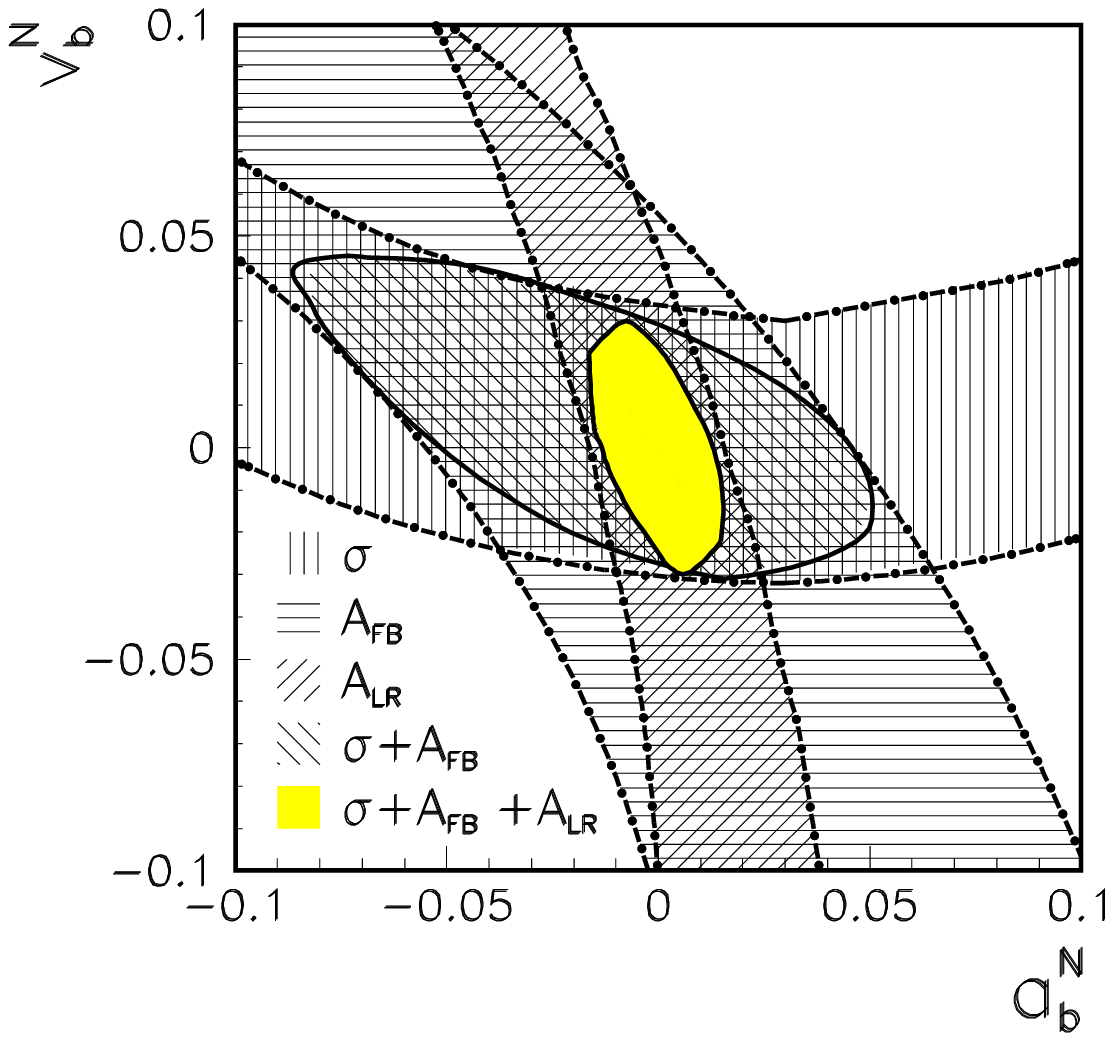}
}
\end{center}
\noindent {\small\bf Fig.~6a: }{\small\it 
Areas in the $(a_b^N,\ v_b^N)$ plane indistinguishable from the SM 
(95\% CL) based on several $b$-quark observables at LEP\,2
and LEP\,2P. 
The leptonic $Z'$ couplings are $a_l^N$=0, $v_l^N$=0.012.
}}\end{minipage}
\hspace{0.5cm}
\begin{minipage}[t]{7.8cm}{
\begin{center}
\mbox{ \epsfysize=7.8cm \epsffile{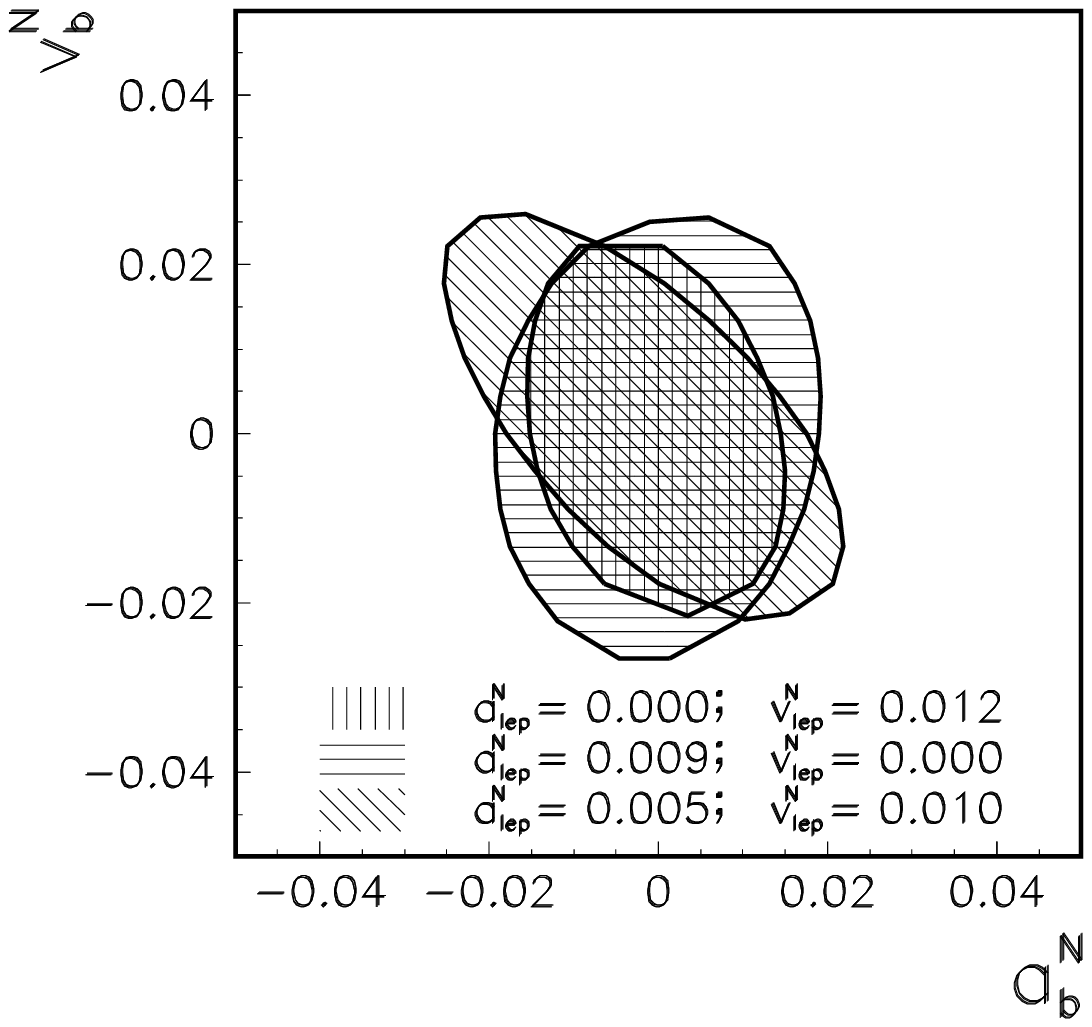}
}
\end{center}
\noindent {\small\bf Fig.~6b: }{\small\it 
Resolution power of LC500 in the $(a_b^N,\ v_b^N)$ plane (95\% CL)
based on a combination of all $b$-quark observables.
Three cases of leptonic $Z'$ couplings on the boundary of fig.~3b are assumed.
}} \end{minipage}
\end{figure}
%

Let us assume  a $Z'$ signal is detected in leptonic observables.
Further, we suppose that the $Z'$ has a 
mass  $M_{Z'}=1.5$\,TeV and is described by one of the models $\chi$, LR or SSM
specified in chapter 1.
As shown in figure~7, in all these cases one can
limit the $Z'b\bar b$ couplings to an area in the $a_b^N, v_b^N$ plane 
around the couplings of the specified model.
However, it is nearly impossible to discriminate between
the $Z'b\bar b$ couplings in the $\chi$ and in the LR models as it could 
be done for $Z'l\bar l$ couplings in figure~4b.

\begin{figure}
\begin{center}
\begin{minipage}[t]{7.8cm} {
\begin{center}
\mbox{ \epsfysize=7.8cm \epsffile{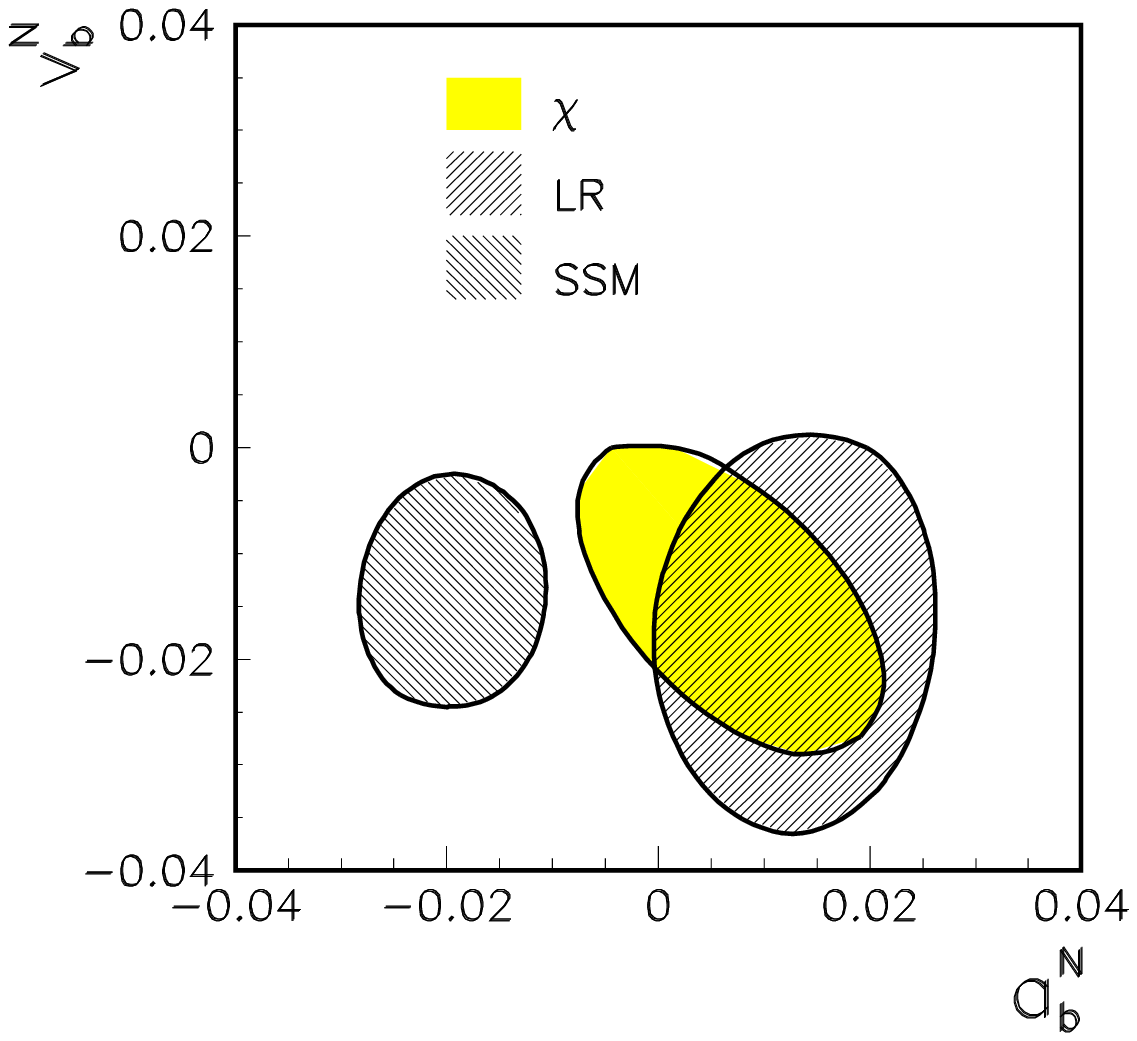}
}
\end{center}
}\end{minipage}
\end{center}
\noindent {\small\bf Fig.~7: }{\small\it 
Expected resolution power of an LC500 in the $(a_b^N,v_b^N)$ plane (95\% CL)
based on a combination of all $b$-quark observables.
Different $Z'$ models are considered, $M_{Z'}$=1.5~TeV. }
\end{figure}
%

%
\section{Model-dependent $Z'$ bounds}

In contrast to the previous sections, we now examine  bounds on
$Z'$ couplings for {\it fixed} $M_{Z'}$ on the one hand
and bounds on $M_{Z'}$ for
{\it fixed} $Z'$ couplings on the other hand. For these studies
we include  those observables listed in
(\ref{mdobs}). 

\subsection{Bounds on $Z'$ couplings}

We follow the method suggested in \cite{delaguila} and define
 certain combinations of leptonic and quarkonic couplings
in order to distinguish between models,
\bq
P_V^l=v_l^N/a_l^N,\ \ P_L^b=(v_b^N+a_b^N)/(2a_l^N)\ \ {\rm and}\ \ 
P_R^b=(v_b^N-a_b^N)/(a_l^N+v_l^N).
\eq
The resolution power of these parameters was tested in \cite{delaguila}
for the models $\chi, \psi, \eta$ and LR. Only statistical errors were included
into the quantitativ considerations there.
We performed the same search
for 1$\sigma$ bounds $(\chi^2<\chi_{min}^2 + 1)$ for $P_V^l,  P_L^b,P_R^b$
and considered additionally
the influence of systematic errors. 
The results are shown in  table~2 for $M_{Z'} = 1$~TeV.
The central values of the parameters $P_V^l,  P_L^b,P_R^b$ are defined
by the $Z'$ model and have to be reproduced in a fit.
The errors of the parameters are in reasonable agreement with \cite{delaguila} 
if systematic errors are neglected. Due to the systematic errors,
the estimated errors of $P_V^l,  P_L^b,$ and $P_R^b$ can increase up to a factor 4.
For $M_{Z'} > 1$~TeV the errors of the fit become larger and
for $M_{Z'} > 2$~TeV the resolution power of $P_V^l,\ P_L^b$ and $P_R^b$ 
is lost completely.

\begin{table}[htb]
\begin{tabular}{lrrrrrr}\hline\\
   & $\chi$ & $\psi$ & $\eta$ & LR \\ \hline\\[0.5ex]
$P_V^l$, no syst. err.   & 2.00$\pm$ 0.11         &
                          0.00$\pm$ 0.064        &
                         -3.00$^{+0.53}_{-0.85}$ &
                         -0.148$^{+0.020}_{-0.024}$ \\
$P_V^l$, syst. err. included & 2.00$\pm$ 0.15 &
                          0.00$\pm$ 0.13 &
                         -3.00$^{+0.73}_{-1.55}$  & 
                         -0.148$^{+0.023}_{-0.026}$  \\
$P_L^b$, no syst. err.   & -0.500$\pm$ 0.018 &
                           0.500$\pm$ 0.035 &
                           2.00$^{+0.33}_{-0.31}$ &
                           -0.143$\pm$0.033  \\
$P_L^b$, syst. err. included & -0.500$\pm$ 0.070 &
                           0.500$\pm$ 0.130 &
                           2.00$^{+0.64}_{-0.62}$ &
                           -0.143$\pm$0.066  \\
$P_R^b$, no syst. err.   &  3.00$^{+0.15}_{-0.14}$ &
                          -1.00$\pm$ 0.29   &
                           0.50$\pm$ 0.11  &
                           8.0$^{+2.5}_{-1.5}$      \\
$P_R^b$, syst. err. included &  3.00$^{+0.65}_{-0.50}$ &
                          -1.00$^{+0.26}_{-0.34}$  &
                           0.50$^{+0.23}_{-0.22} $ &
                          8.0$^{+6.7}_{-2.4}$  \\
                        & & & & \\
\hline
\end{tabular}\medskip

{\small \bf Table~2: }{\small\it  $Z'$ coupling
combinations  $P_V^l,\ P_L^b$ and $P_R^b$
and their 1$\sigma$ errors derived from all observables with and without
systematic errors for $\sqrt{s}$=500~GeV and  $M_{Z'}$=1~TeV.
} \end{table}
%

\subsection{Bounds on  $M_{Z'}$ }

We now search for lower limits (95\% CL; $\chi^2 < \chi^2_{min} + 2.7)$
for  $M_{Z'}$ in different $Z'$ models.
The expected lower $Z'$ mass limits, $M_{Z'}^{lim}$,
are listed in table~3, which is subdivided into  three rows
in accordance with different collider  scenarios.
The first row gives $M_{Z'}^{lim}$
 based on leptonic observables only.
The second (third)
row shows the results of an analysis including
all leptonic and hadronic observables with (without) systematic errors.
The numbers for LC2000 are obtained under the assumption
that the radiative corrections considered in the available programs
work up to these energies.
Comparing the first two rows for a certain collider, we see that the
hadronic observables improve the mass limits by 5\% to 10\%.
One may conclude that leptonic and hadronic observables
are approximately equally important for the determination of $M_{Z'}^{lim}$.
The LR and the SSM are an exception. Their mass limits are 
mainly determined by hadronic observables.
But for hadronic observables the systematic error is large with respect to
the expected statistical uncertainty. 
Hence, an analysis neglecting systematic errors
suggests relatively high mass bounds for these models.
In the case of the other models, the $Z'$ mass limits
are rather insensitive to the anticipated systematic errors.
Neglecting them, one gets an overestimation of $M_{Z'}^{lim}$ 
by approximately 10\%.

We conclude that $e^+e^-$ colliders can either detect a $Z'$ or exclude a
$Z'$ with a mass less than $3$ to $6 \sqrt{s}$
for typical GUTs and  up to  $8\sqrt{s}$ for the SSM.
Furthermore, we see that the $Z'$ limits from LEP\,2 can compete with
the limits expected from the Tevatron \cite{godfrey}.

\begin{table}
\begin{tabular}{lrrrrrr}\hline\\
  & & $\chi$ & $\psi$ & $\eta$ & LR & SSM\\ \hline\\[0.5ex] 
LEP\,2    &leptonic observables only &  890 &  490 &  540 &  680 &  940 \\
          &all observables           &  940 &  540 &  580 &  760 & 1500 \\
&all observables no systematic errors& 1000 &  610 &  650 &  860 & 1700 \\[2ex]
LEP\,2P   &leptonic observables only &  940 &  490 &  560 &  690 &  950 \\
          &all observables           &  990 &  560 &  620 & 1100 & 1500 \\
&all observables no systematic errors& 1100 &  640 &  690 & 1300 & 1700 \\[2ex]
LC500    &leptonic observables only & 2600 & 1400 & 1600 & 2000 & 2700  \\
          &all observables           & 2800 & 1600 & 1700 & 3200 & 4000 \\
&all observables no systematic errors& 3100 & 1800 & 1900 & 3800 & 4700 \\[2ex]
LC2000   &leptonic observables only &10000 & 5800 & 6400 & 8200 &11000  \\
          &all observables           &11000 & 6300 & 7000 & 9300 &17000 \\
&all observables no systematic errors&13000 &  7300& 7700 &10800 &19000 \\
\multicolumn{2}{l}
{Tevatron}  & & & & &\\
\multicolumn{2}{l}
{($\sqrt{s}=1\,TeV,\ L_{int}=1fb^{-1}$)} &
 775 & 775 & 795 & 825 &  \\ 
\hline
\end{tabular}\medskip

{\small\bf Table~3: }{\small\it Lower $Z'$ mass bounds (95\% CL),
$M_{Z'}^{lim}$ in GeV, derived for  LEP\,2, LC500 and  LC2000.
The Tevatron bounds are  from \cite{cg,godfrey}.
} \end{table}

\section{Conclusions}

We performed a model-independent $Z'$ analysis for LEP\,2 and for $e^+e^-$
colliders with centre-of-mass energies of 500\,GeV and 2\,TeV.
We compared different processes and found that fermion pair production
is most sensitive to potential $Z'$ contributions.
We took into account all available radiative corrections and
expected statistical and systematic errors.
Observables measured at LEP\,1 are assumed to be measurable also at
higher energies.
Popular $Z'$ models are discussed as special cases of the
model-independent approach.
For a given model, the ratio $M_{Z'}^{lim}/\sqrt{s}$ is almost
constant for the considered collider scenarios. A $Z'$ predicted by
usual GUTs could be detected if its mass is less than 6$\sqrt{s}$.
The resolution power between different models is studied for the case
of a $Z'$ signal. A reasonable model discrimination is shown
to be feasible if the $Z'$ is lighter than $M_{Z'}^{lim}/2$.

While polarized beams give only a minor improvement to {\it exclusion} limits,
they are quite important for {\it measurements} of the $Z'$ couplings to fermions.

Systematic errors have only a slight influence on the {\it exclusion} bounds
for the $Z'$ mass while
{\it measurements} of the $Z'$ couplings to fermions
are very sensitive to them.

Calculations in Born approximation are sufficient for theoretical
predictions of potential $Z'$ limits. For fits to future data
radiative corrections, kinematical cuts und the inclusion of
systematic errors are essential.
With the existing program {\tt ZEFIT}, which works with {\tt ZFITTER}
a comprehensive $Z'$ analysis of LEP\,2 data can be performed.

%
%
\ \vspace{1cm}\\ {\Large\bf Acknowledgement\hfill}\vspace{0.5cm}\\
We would like to thank C. Verzegnassi for stimulating this work
and T. Riemann for a careful reading of the manuscript.
\\ \\
 {\Large\bf Note added:}\\ \\
After submission of the paper we discovered that
the numbers in table~3 correspond to an older version of
our study. We thank A.A. Pankov and N. Paver who directed our attention
to that. Compared to the preprint DESY 96--111; LMU--96/02
table~3 has been corrected here.

\end{document}